\documentclass[aps,pra,twocolumn]{revtex4-1}

\usepackage{amsmath}
\usepackage{amssymb}

\usepackage{xcolor}

\usepackage{setspace}
\usepackage{bm}
\usepackage{graphicx}
\numberwithin{equation}{section}

\begin{document}


\title{Active Cancellation of Acoustical Resonances with an FPGA FIR Filter}
\author{Albert Ryou \& Jonathan Simon}
\affiliation{James Franck Institute and the Department of Physics at the University of Chicago}
\date{\today}

\begin{abstract}
We present a novel approach to enhancing the bandwidth of a feedback-controlled mechanical system by digitally canceling acoustical resonances (poles) and anti-resonances (zeros) in the open-loop response via an FPGA FIR filter.  By performing a real-time convolution of the feedback error signal with an inverse filter, we can suppress arbitrarily many poles and zeros below 100 kHz, each with a linewidth down to 10 Hz. We demonstrate the efficacy of this technique by canceling the ten largest mechanical resonances and anti-resonances of a high-finesse optical resonator, thereby enhancing the unity gain frequency by more than an order of magnitude. This approach is applicable to a broad array of stabilization problems including optical resonators, external cavity diode lasers, and scanning tunneling microscopes, and points the way to applying modern optimal control techniques to intricate linear acoustical systems.
\end{abstract}

\maketitle

\section{\label{sec:Intro} Introduction}
Active stabilization is a crucial tool for applications spanning physics and engineering. Lasers, resonators, interferometers, clocks, and even noise-cancelling headphones employ closed-loop feedback for noise suppression and real-time control. In the case of noise suppression the ultimate limit is set by the feedback control system's ``reaction time'': the delay between when noise in a system is first detected and when that system physically responds \cite{bechhoefer2005feedback}. This determines the range of frequencies that the feedback can suppress, and thus limits the unity gain frequency (or bandwidth) of the control path.

In controlling mechanical systems, one rarely reaches the time-delay limited bandwidth, due to parasitic coupling to acoustical resonances. These low-frequency vibrational modes can range in frequency from a few Hz to several hundred kHz, and attempts to suppress noise at or beyond these resonance frequencies result in spontaneous oscillation of the system due to positive feedback. The most widely used control mechanism, the proportional-integral-derivative (PID) controller, can cancel a single acoustical resonance, but in practice the complexity involved with fine-tuning to achieve precise cancellation, along with the necessity of rolling off the derivative gain at high frequency, leads many experimenters to omit derivative feedback and accept lower bandwidth \cite{bechhoefer2005feedback}.

The bandwidth limitation imposed by the resonances have led a number of groups to modify the mechanical structure itself to either damp the resonances or push them to higher frequencies. In stabilizing a femtosecond laser cavity with piezoelectric transducer (PZT) fiber-stretchers, Sinclair et al. damp numerous ``violin-like'' resonances between 1 kHz and 100 kHz with modeling clay and electrical tape \cite{sinclair2015invited}. Chadi et al. employ a side-clamping holder for their piezo-actuator \cite{chadi2013note}, utilizing structural symmetry for enhanced common-mode rejection of longitudinal coupling, an idea that has been used to great effect for laser stabilization \cite{notcutt2005simple}. Briles et al. glue their actuator to a tapered copper mount that is filled with lead \cite{briles2010simple}. These methods, while highly effective, impose severe design constraints.

A more flexible approach is to modify the control law with inversion-based optimal control: in this approach, the set-point is filtered with the inverse of the system's dynamic response to avoid exciting mechanical resonances \cite{bechhoefer2005feedback}. This technique has been successfully employed to accelerate step-response outside of the loop \cite{singer1997input, croft1999vibration, schitter2004identification, zou2005precision, ha2013robust}, but has yet to be applied to in-loop noise suppression. This is because improving the step response is not constrained by latency, and maybe pre-computed offline. By contrast, noise suppression requires real-time, low-latency loop shaping.
   
With recent advances in field-programmable gate arrays (FPGAs), low-latency digital feedback has become a viable alternative to its analog counterpart, offering enhanced tunability a loop control. Yang et al., Schwettmann et al., and Sparkes et al. implement FPGA-based PID controllers \cite{yang2012low, schwettmann2011field, sparkes2011scalable}, and Leibrandt et al. manage to configure a notch filter to cancel the lowest-frequency mechanical resonance of a doubling cavity \cite{leibrandt2015open}. Compensating more complex mechanical systems requires sophisticated loop shaping that may be digitally implemented with finite impulse response (FIR) or infinite impulse response (IIR) filters. While such  FPGA-based filters have been demonstrated by a number of groups \cite{do1998flexible, evans1994efficient, chou1993fpga, yoo2005hardware}, they have not been applied to noise suppression in acoustical systems.


In this paper, we present a novel digital architecture for loop-shaping in acoustical feedback systems. Harnessing the massive processing power of a state-of-the-art FPGA, we demonstrate a low-latency, 25,600-tap FIR filter capable of \emph{precisely} canceling an arbitrary number of acoustical resonances (poles) and anti-resonances (zeros) below $\sim$100 kHz, thereby enhancing the noise-suppression bandwidth by more than an order of magnitude.


In Section \ref{sec:Theory}, we summarize single-input-single-output feedback control, as it relates to the system transfer function, the limits imposed by causality and time delay, and the FPGA FIR filter. In Section \ref{sec:Experiment}, we provide a step-by-step illustration of our technique: we measure the transfer function of a high-finesse optical resonator locked to an external reference laser, generate an inverse filter, and demonstrate more than an order of magnitude increase in feedback bandwidth. Finally, in Section \ref{sec:Conclusion} we conclude by exploring potential applications in areas beyond atomic physics.


\section{\label{sec:Theory} Principles of operation}

\subsection{Feedback bandwidth and stability}
\begin{figure}
\includegraphics[width=86mm]{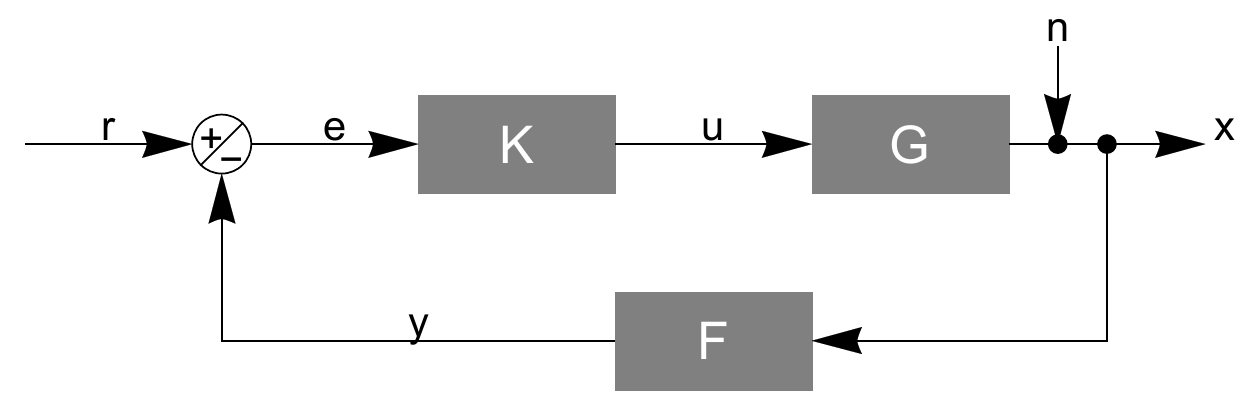}
\caption{\label{Figure:blockdiagram} \textbf{Block diagram of a general feedback loop}. The output from the detector in the physical system $G$, is first fed into the FPGA-based digital FIR filter $F$, which conditions it to remove the acoustical resonances and anti-resonances. Finally, the feedback controller $K$ (typically proportional + integral gain), takes the difference between this signal and the reference $r$ and feeds it back into the system to stabilize it.}
\end{figure}

Figure \ref{Figure:blockdiagram} shows the block diagram of a generic feedback loop. The block term $G$ denotes the physical system to be controlled, $K$ denotes the controller, and $F$ denotes the FPGA.  The signals $r$, $e$, $u$, $n$, $x$, and $y$ denote the reference, the error, the system input, the noise, the system output, and the FPGA-modified output, respectively.

In the frequency domain, the effect of the noise on the output can be calculated as follows:
\begin{align}
x &= KGe + n \\
y & = Fx \\
e &= r - y \\
\label{eqn:loopeqns}
x &= \frac{KG}{1+KGF} r + \frac{1}{1+KGF} n
\end{align}

The feedback suppresses noise by a factor $1/(1+KGF)$ called the sensitivity \cite{bechhoefer2005feedback}. The typical behavior of the sensitivity is decreasing suppression of the noise up to the unity gain frequency, where the magnitude of the loop gain $|KGF| = 1$. We define the feedback bandwidth to be equal to the unity gain frequency. In practice, the bandwidth is principally controlled by adjusting a multiplicative pre-factor inside of $K$ that we will call the ``total gain''.

\begin{figure*}
\includegraphics[width=178mm]{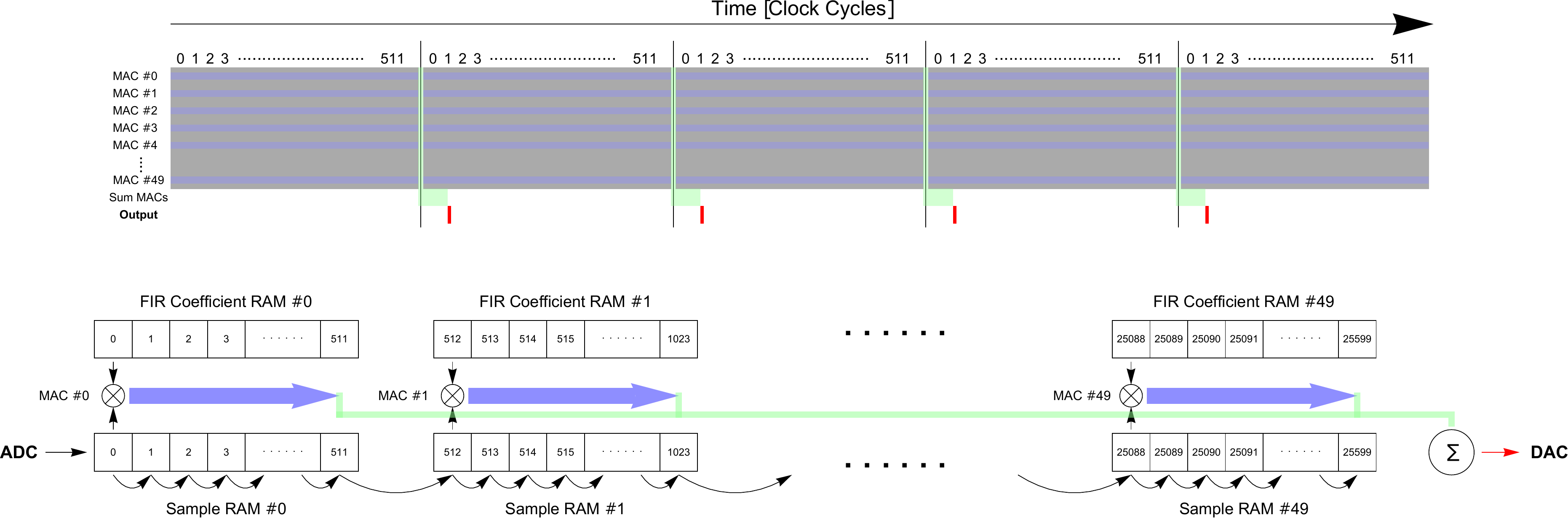}
\caption{\label{Figure:FPGA_fir_impl} \textbf{FPGA FIR filter implementation}. The FIR filter is a real-time digital implementation of a temporal impulse response function. The last 25,600 values of a sampled input signal from an analog to digital converter (ADC) are multiplied by 25,600 user-defined filter coefficients, and the results are summed and sent to a digital to analog converter (DAC). This calculation is performed by 50 Multiply-ACcumulators (MACs) in parallel, each carrying out $512$ multiplications and additions in series, which are then summed in series. (top) Timing diagram for the series-parallel FIR filter implementation in the FPGA. The $\sim512$ clock cycles of each sample period are broken down, reflecting when the 50 parallel MACs are operating, and when their results are summed in series and output. (bottom) Connectivity diagram of the implementation, reflecting the 50 sample and coefficient RAMs, each connected to a MAC, and the path an input sample takes through the sample RAMs over time.}  
\end{figure*}

As can be seen from Equation (\ref{eqn:loopeqns}), when $KGF = -1$ the feedback becomes unstable, leading to spontaneous oscillation. Thus, the bandwidth cannot be increased beyond the frequency at which the phase of the loop gain approaches $\pi$. Acoustical (anti-)resonances cause the phase and amplitude of $KGF$ to vary violently within a small frequency range, ensuring that unless the total gain of the feedback path is reduced aggressively, (a) the open-loop feedback amplitude will go through unity, and (b) the phase will be very nearly $\pi$, inducing oscillations. We will thus need to quantitatively characterize the most general frequency response and suppress the resulting resonances.

\subsection{Character of the physical system $G$}
The transfer function of any linear system without a time delay (i.e. minimum phase \cite{bechhoefer2005feedback}) can be written as a rational function of the frequency $\omega$:
\begin{equation}
G(\omega) = \frac{M(\omega)}{N(\omega)}
\end{equation}
where $M(\omega)$ and $N(\omega)$ are polynomials whose roots are called zeros and poles, respectively, of the system response. For a general physical system, the polynomials may be arbitrary as long as the corresponding impulse response function (time-domain Green's function, which is the inverse Fourier transform of the frequency response) is both real and causal.

Realness of the time-domain impulse response implies that $G^*(\omega) = G(-\omega)$ for real $\omega$ \cite{dorf1992modern}; thus, all poles and zeros of the Fourier-domain transfer function either occur on the imaginary axis, or come in pairs reflected across the imaginary axis. To satisfy causality, the poles and zeros of a stable system must be located in the upper half of the complex $\omega$-plane such that a contour integration yields a Green's function that is zero for $t<0$ and nonzero for $t \ge 0$ \cite{dorf1992modern}.

Building upon these ideas, we factorize the transfer function of the physical system $G$ as:
\begin{equation}
\label{eqn:model}
G(\omega) = A e^{i\omega \tau_G}          \left[         \frac{   \prod_j \left( \omega - i\gamma_z^j \right)  }   {\prod_j \left( \omega - i\gamma_p^j \right)     }              \right]        \left[     \frac{   \prod_k \left( \omega - \omega_z^k \right)     \left( \omega + \omega_z^{k*} \right)}   {\prod_k \left( \omega - \omega_p^k \right)    \left( \omega + \omega_p^{k*} \right)  }  \right]
\end{equation}
where $A$ is a real amplitude, $\tau_G$ is a time delay, the products over $j$ are over first-order zeros and poles, $\gamma_{z,p}$, and the products over $k$ are over second-order zeros and poles, $\omega_{z,p}$. The latter are complex frequencies whose imaginary part reflects the linewidth of the corresponding pole or zero.

Given the system $G$ and the controller $K$, the role of the FPGA is to implement a filter response $F$ such that the loop gain $KGF$ exhibits a smooth $1/\omega$ behavior whenever $|KGF|$ is within an octave of unity; this ensures that the phase of $KGF$ is approximately $\pi/2$, and thus that the sensitivity $1/(1+KGF)$ does not diverge.


\subsection{FPGA FIR filter}
An FIR filter is a digital implementation of a discrete, time-domain Green's function \cite{oppenheim1989discrete}. A number of samples are convolved with an equal number of filter coefficients to produce a filter with compact support in time (hence \emph{finite} impulse response, FIR). The FIR filter and its Z-transform \cite{oppenheim1989discrete} are given by:
\begin{align}
y_n &= \sum_{m=0}^{J_{max}-1} a_m x_{n-m} \\
\tilde{y} &= \left[ \sum_m a_m z^m \right] \tilde{x}
\end{align} 
where $z=e^{i\omega \tau_s}$ and $\tau_s = \frac{1}{f_s}$ is the sampling period, $x_n$ is the input sample at time $t=n\times\tau_s$ and $y_n$ is the resulting output sample. $\tau_s$ determines the maximum frequency of the filter (1/(2$\tau_s$)), and the number of samples included in the filter $J_{max}$ determines the spectral resolution (1/($\tau_s J_{max})$), and hence quality factor, of the poles and zeros of the filter. This filter is the time-domain implementation of an arbitrary Green's function $F(t)\approx\sum_{m=0}^{J_{max}-1}a_m Sq(t-m\tau_s)$, where $Sq(t)$ equals one if $t\in[0,\tau_s]$ and zero otherwise. Note that this implementation already \emph{implies} a system time delay of $\tau_s/2$.

We employ a filter with $J_{max} =$ 25,600 taps (coefficients, 17-bit). The filter is implemented using $N_{MAC} = 50$ parallel multiply-accumulators (MACs) (realized with DSP48 slices in the FPGA), each capable of carrying out $n_{op}=512$ serial multiplications and additions within one sample period. This implementation yields a sampling rate $f_s = 243$ kHz $\approx \frac{f_{clock}}{n_{op}}$ and a delay of $\tau = 2.6$ $\mu$s $\approx\frac{1}{2 f_s} + \frac{N_{MAC}}{f_{clock}}$, where the FPGA's clock rate is $f_{clock} = 125$ MHz. See Figure \ref{Figure:FPGA_fir_impl} for a diagram of the FIR implementation. 

While an equivalent infinite impulse response (IIR) filter requires only as many taps as the number of poles and zeros in the desired filter function, each MAC operation must be performed to much higher accuracy to avoid numerical instability; on the other hand, the FIR implementation requires a huge number of taps to accurately simulate the poles, but with less required accuracy in the MAC operations \cite{oppenheim1989discrete}. It is only the massively-parallel nature of the FPGA that makes such high-throughput FIR architectures possible.

The reconfigurable logic employed in this work is a commercial Red Pitaya board (redpitaya.com, $\sim$\$240 at the date of publication) that contains a Xilinx Zynq 7010 SoC consisting of a Dual core ARM Cortex A9+ processor running network-enabled Linux as well as a 28k logic cell Artix-7 FPGA with 80 DSP48 slices and 2.1Mb BRAM, and 2 channel, 14-bit DAC/ADCs capable of sampling at $>$100MSPS. The FGPA's configuration is determined by a custom Verilog code implementing an FIR filter; the filter's discrete impulse response (coefficients $a_m$) is calculated offline via Python. 

\subsection{Anti-aliasing filter}
The discrete sampling of a continuous-time system output $x(t)$ at $f_s = 243$ kHz by the FPGA leads to aliasing: noise above the Nyquist frequency ($f_{Nyquist} = f_s/2$) is aliased into the base band. We place before the ADC input of the FPGA an anti-aliasing low-pass filter whose corner frequency, $\omega_{corner} = 2\pi \times 100$ kHz, has been chosen to trade-off between suppressing aliased noise and the phase lag induced by the anti-aliasing filter itself.


\section{\label{sec:Experiment}Application:\\Canceling acoustical resonances of an optical resonator}

\subsection{Description of the optical resonator}
\begin{figure}
\includegraphics[width=86mm]{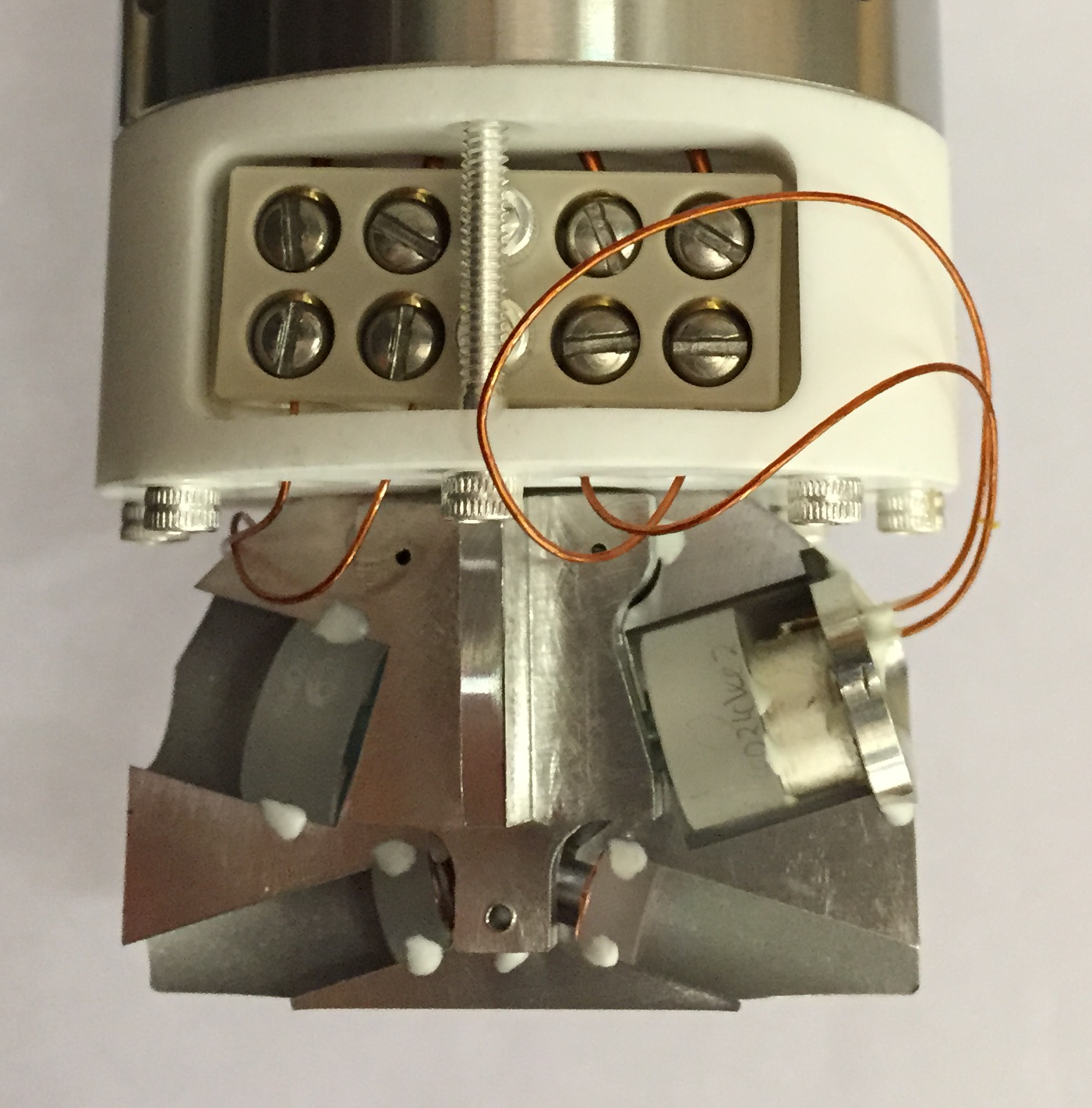}
\caption{\label{Figure:resonator} \textbf{Experimental optical resonator}. The resonator, shown here outside a vacuum chamber, consists of four mirrors in a bow-tie, running-wave configuration, one of which is mounted on a single-layer piezoelectric tube actuator. Because the resonator has been designed for a small waist, degenerate optical modes, and large mirror separation, instead of mechanical rigidity or vibration isolation, it exhibits numerous low-frequency acoustical resonances that couple to the piezo-actuator, making this an ideal candidate for control using the digital FIR filter.}
\end{figure}

We apply the FIR filter to canceling the acoustical resonances present in a high-finesse optical resonator designed for Rydberg cQED experiments \cite{rydbergPaper}. In order to have a small waist size of about 10 $\mu$m, comparable to Rydberg blockade, the resonator consists of four mirrors in a bow-tie configuration, one of which is mounted on a single-layer piezoelectric tube actuator. The piezo-actuator is affixed to a 2-mm thick stainless-steel wafer, which is itself glued to the main spacer. A separate stainless steel piece encloses the mode waist to suppress stray electric fields. All of these components, plus vacuum wires, give rise to numerous low-frequency acoustical resonances that couple to the actuator. See Figure \ref{Figure:resonator} for a photograph of the resonator.           

The purpose of the piezo-actuator is to tune the length of the resonator, which is frequency-modulation locked (the transmission analog of the Pound-Drever-Hall method \cite{drever1983laser, black2001introduction}) to a stable external laser. The feedback controller is an analog proportional-integral (PI) controller, whose output is amplified by a high-voltage piezo-driver before reaching the piezo-actuator.

\begin{figure}
\includegraphics[width=86mm]{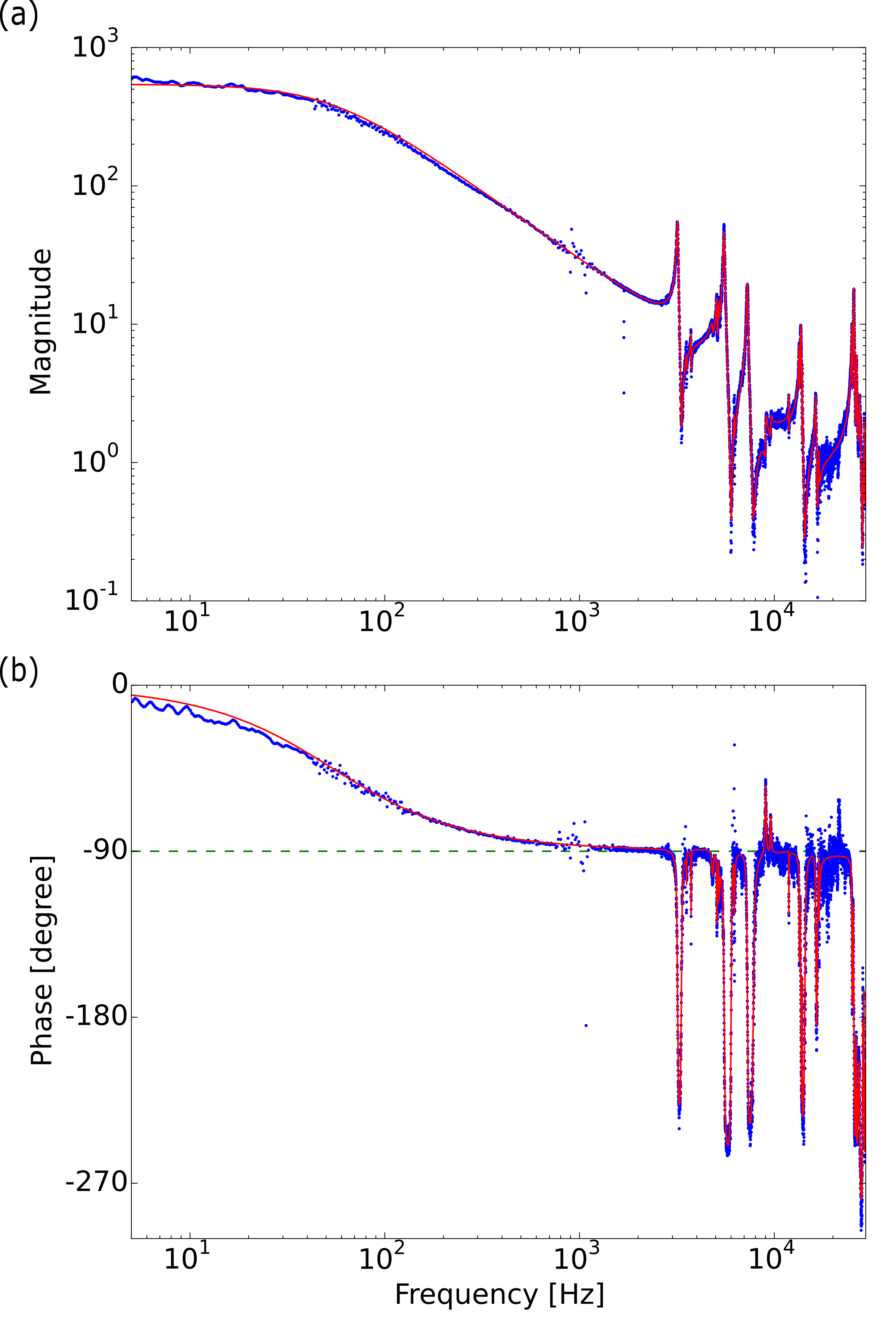}
\caption{\label{Figure:standalone} \textbf{System transfer function $G$}. (a) Magnitude and (b) Phase of the system transfer function, versus Frequency. Here ``system'' refers to the piezo-driver, piezo-actuator, resonator mirror, photodiode, and down-converting mixer. The measured data (blue) agrees well with a fit (red) composed of a rational function of poles and zeros. The first-order imaginary pole (corner frequency $\omega_{piezo} = 2\pi \times 50$ Hz) is due to the piezo-actuator-piezo-driver circuit; its $\pi/2$ phase lag is marked with a dashed line in (b). The second-order poles and zeros, appearing in pairs, are the result of constructive and destructive interference between the piezo-actuator and mount modes. The phase lag within each pole-zero pair depends on the strength of the coupling between the piezo and the mount, and can be as large as $\pi$. The fit to the magnitude data in (a) provides a predicted phase fit in (b) that agrees well with the measure phase, indicating that the system exhibits negligible time delay over the measured frequency range.}
\end{figure}

\subsection{Measurement of the system transfer function $G$}
To measure the transfer function $G$ of the full system comprised of the piezo-driver, piezo-actuator, mirror, resonator field, and detector, we modulate the reference signal $r$ with a network analyzer (HP3577A) and simultaneously record the ratio detector output to the piezo-driver input $u$, as a function of frequency. For strong modulation, $n \ll x$, and so $G = x/u$. 

Figure \ref{Figure:standalone} shows the magnitude and the phase of $G$, measured from 5 Hz to 30 kHz. We observe the first-order pole, coming from the output impedance of the piezo-driver ($\sim$1 M$\Omega$) and the capacitance of the piezo-actuator ($\sim$4 nF), with a corner frequency of $\omega_{piezo} = 2\pi \times 50$ Hz. As expected, the phase lag of the first-order pole is $\pi/2$, marked by the dashed line in Figure \ref{Figure:standalone}b. Past the corner frequency is a $1/f$ decline, on top of which the mechanical resonances and the anti-resonances appear one after the other.

The resonances and the anti-resonances are second-order poles and zeros, which come with a phase shift of $-\pi$ and $\pi$, respectively. Experimentally, they occur in pairs, arising from the interference between the motion of the piezo and that of other mechanical parts of the resonator (the ``mount''): the motion of the piezo excites mount modes, which can back-act on, and hence interfere with, the motion of the piezo itself. Constructive interference results in a pole, while complete destructive interference causes a zero. Within the pole-zero pair, the phase is nearly $\pi$ or $-\pi$, depending on the order in which they occur. If the pole and the zero are too close together (if the mount mode is narrow and weakly coupled to the piezo), the phase lag between them is smaller than $\pi$. See supplementary information for a physical model of the resonances.

\subsection{Extraction of the inverse filter parameters}
We extract the frequencies and the linewidths of the acoustical resonances by fitting the measured system transfer function $G$ to the minimum-phase mathematical model, Equation (\ref{eqn:model}), with $\tau_G=0$.  Figure \ref{Figure:standalone} shows the result of a least-squares fit to the magnitude, and plots both the magnitude (Fig. \ref{Figure:standalone}a) and the phase (Fig. \ref{Figure:standalone}b) of the fit in red. That the measured and calculated phase agree so well is an indication that the time delay $\tau_G$ in the physical system is negligible over the measured frequency range.

While we can fit as many features below $f_{Nyq}$ as we wish by including an arbitrary number of poles and zeros (in Figure \ref{Figure:standalone}, we fit 28 pole-zero pairs), the most important features when it comes to increasing the feedback bandwidth are the largest poles and zeros. For the FIR filter, we identify and cancel ten pole-zero pairs, treating closely-spaced resonances as single lumps. In Table \ref{tab:resdata}, we list the frequencies and the linewidths of six of the ten pairs, corresponding to those shown in Figure \ref{Figure:standalone}.

\begin{table}
\caption{\label{tab:resdata} Fitted frequencies and linewidths of largest second-order poles and zeros of the system transfer function $G$. Many of these features have quality factors upwards of 200, making analog cancellation prohibitively sensitive.}
\begin{ruledtabular}
\begin{tabular}{ll}
\multicolumn{2}{c}{\textbf{Poles}} \\
\textbf{Frequency [Hz]} & \textbf{FWHM [Hz]} \\
3190 & 30 \\
5530 & 60 \\
7290 & 70 \\
13700 & 110 \\
16350 & 100 \\
25530 & 400
\end{tabular}
\begin{tabular}{ll}
\multicolumn{2}{c}{\textbf{Zeros}} \\
\textbf{Frequency [Hz]} & \textbf{FWHM [Hz]} \\
3330 & 30 \\
6000 & 30 \\
7810 & 80 \\
14350 & 140 \\
16600 & 190 \\
28400 & 40
\end{tabular}
\end{ruledtabular}
\end{table}

\subsection{Implementation of the FIR filter}
\begin{figure}
\includegraphics[width=86mm]{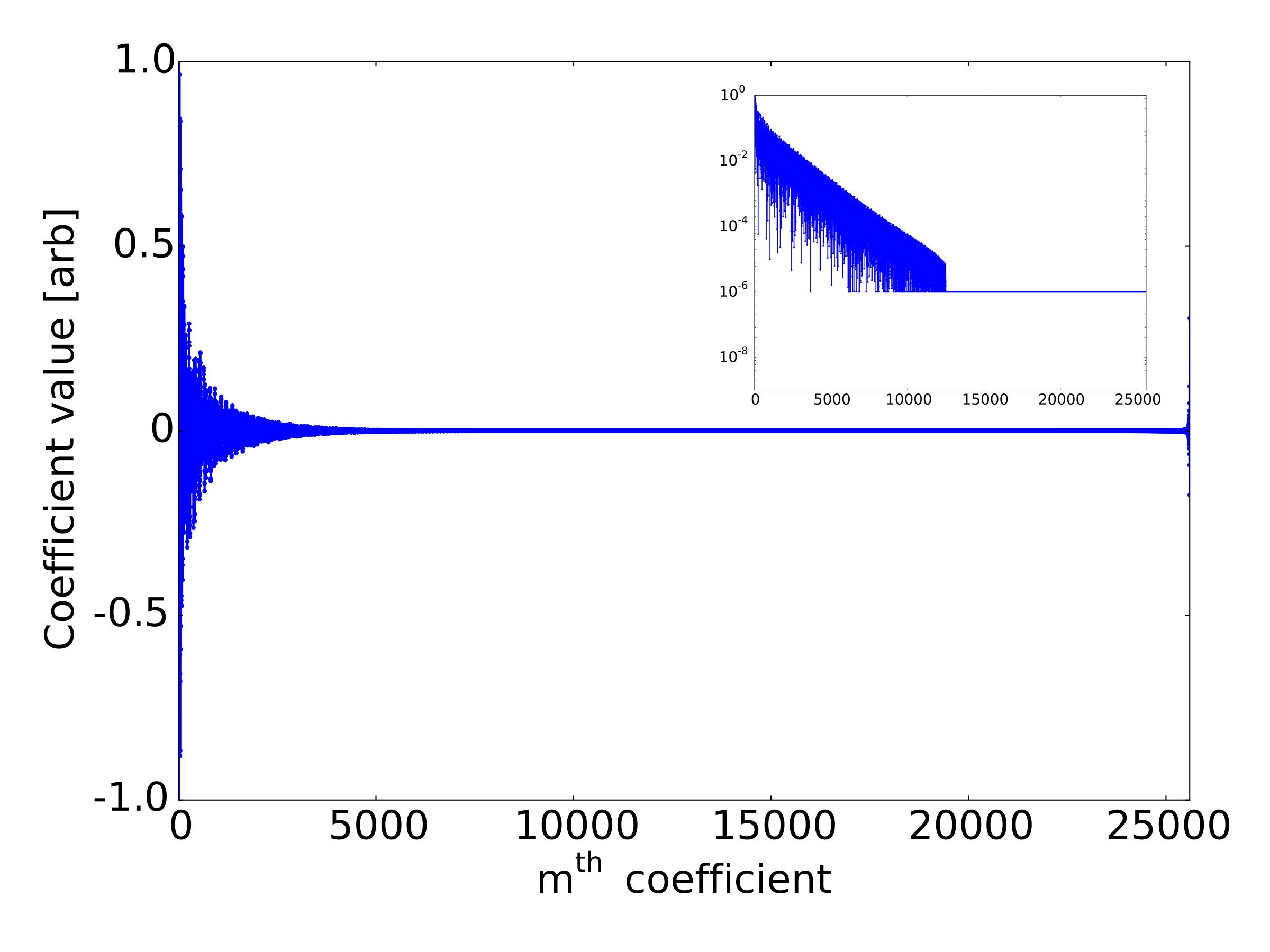}
\caption{\label{Figure:time} \textbf{FIR filter coefficients}. The 25,600 filter coefficients that are convolved with the output signal $x$. (\emph{inset}) Absolute value of the filter coefficients on a log scale; values below quantization error have been truncated. The fact that a large number of coefficients are zero indicates that, for this system, it would be possible to reduce $J_{max}$ to achieve a higher sampling frequency $f_s$, and a correspondingly lower latency $\tau$.}
\end{figure}

The inverse filter $F$ that the FPGA FIR implements is computed by simply exchanging the second-order poles and zeros in $G$, that is, $F(\omega) = [G(\omega)]^{-1}$, excluding the time delay and the first-order pole.  We then generate the filter coefficients $a_m$ by sampling the analytic expression for $F$ at the rate $f_s$ and performing an inverse discrete Fourier transform. Figure \ref{Figure:time} shows the values of the filter coefficients for the inverse filter $F$. The gain of $F$, and hence $KGF$, can be adjusted by changing the normalization of the coefficient values.

\begin{figure}
\includegraphics[width=86mm]{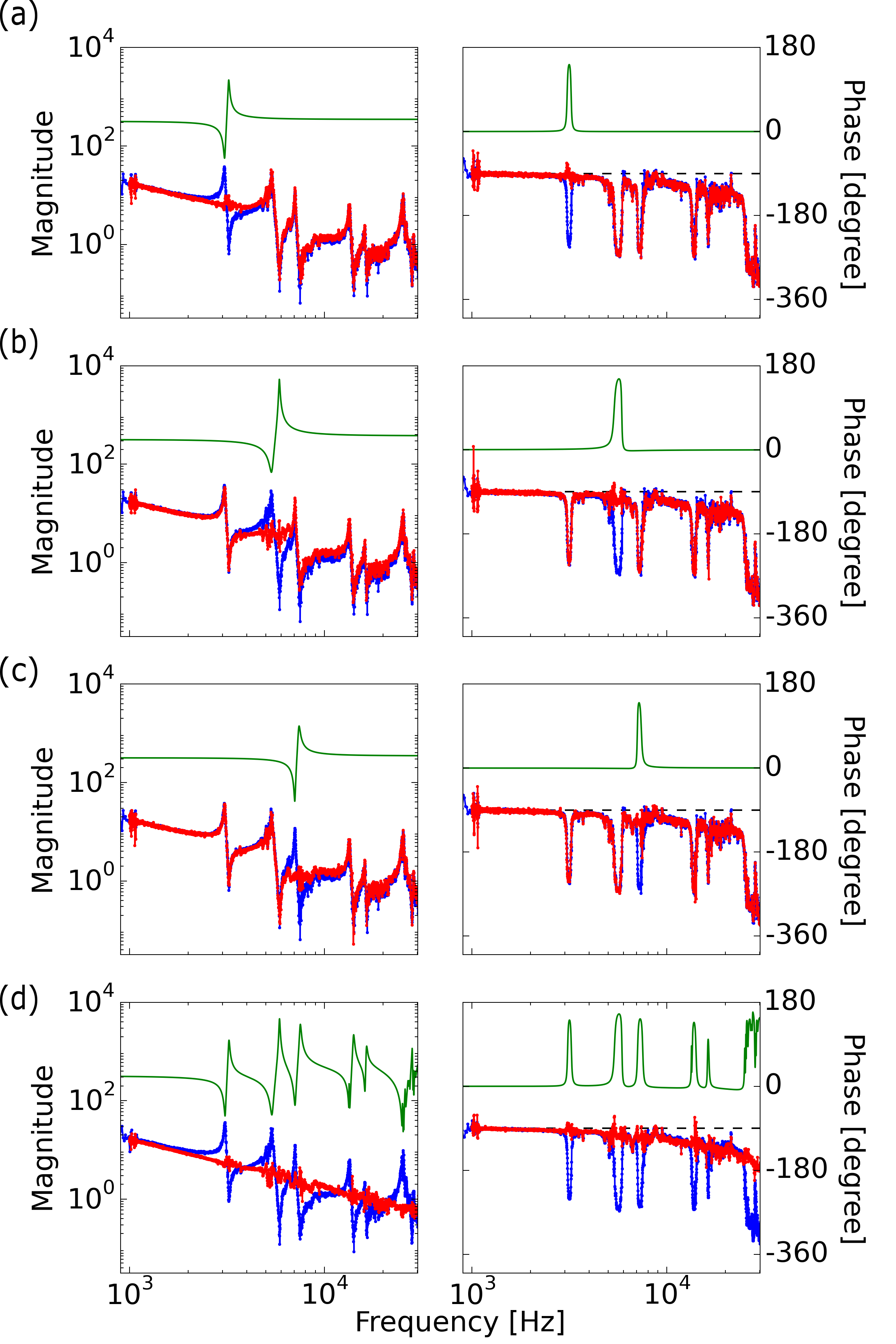}
\caption{\label{Figure:selective} \textbf{Selective acoustical resonance cancellation using the FPGA FIR filter}. (a-c) Each of the first three pairs; (d) all six pairs. The magnitude and the phase of the measured transfer function $GF$ with $F=1$ (blue) and $F = $ inverse filter (red) are plotted. Also shown is the calculated transfer function of the inverse filter $F$ (green), not including the 2.6-$\mu$s time delay in the digital filter; the magnitude of $F$ has been offset from unity for clarity. The time delay results in an increasing phase lag for higher frequency for the measured transfer functions, as shown by the divergence from the horizontal dashed line at $-90$ degrees.}  
\end{figure}

Figure \ref{Figure:selective} demonstrates our ability to cancel select poles and zeros of the transfer function $G$ with the FIR filter. The left column shows the magnitude, and the right the phase, of $GF$. The blue curve reflects measured $GF$ with $F = 1$, and the red curve reflects the measured $GF$, where $F$ is the inverse filter, displayed in green. In Figures \ref{Figure:selective}a-c, we selectively cancel each of the first three strongly-coupled mechanical modes; Figure \ref{Figure:selective}d shows the cancellation of all six features simultaneously.

Unlike the phase of the system transfer function $G$ in Figure \ref{Figure:standalone}, the phase of $GF$ in Figure \ref{Figure:selective} shows an increasing phase lag at higher frequencies, $\phi=\omega \tau$, arising from the $\tau=$ 2.6 $\mu$s time delay caused by the sampling rate of FPGA FIR filter, as described previously.    

\begin{figure}
\includegraphics[width=86mm]{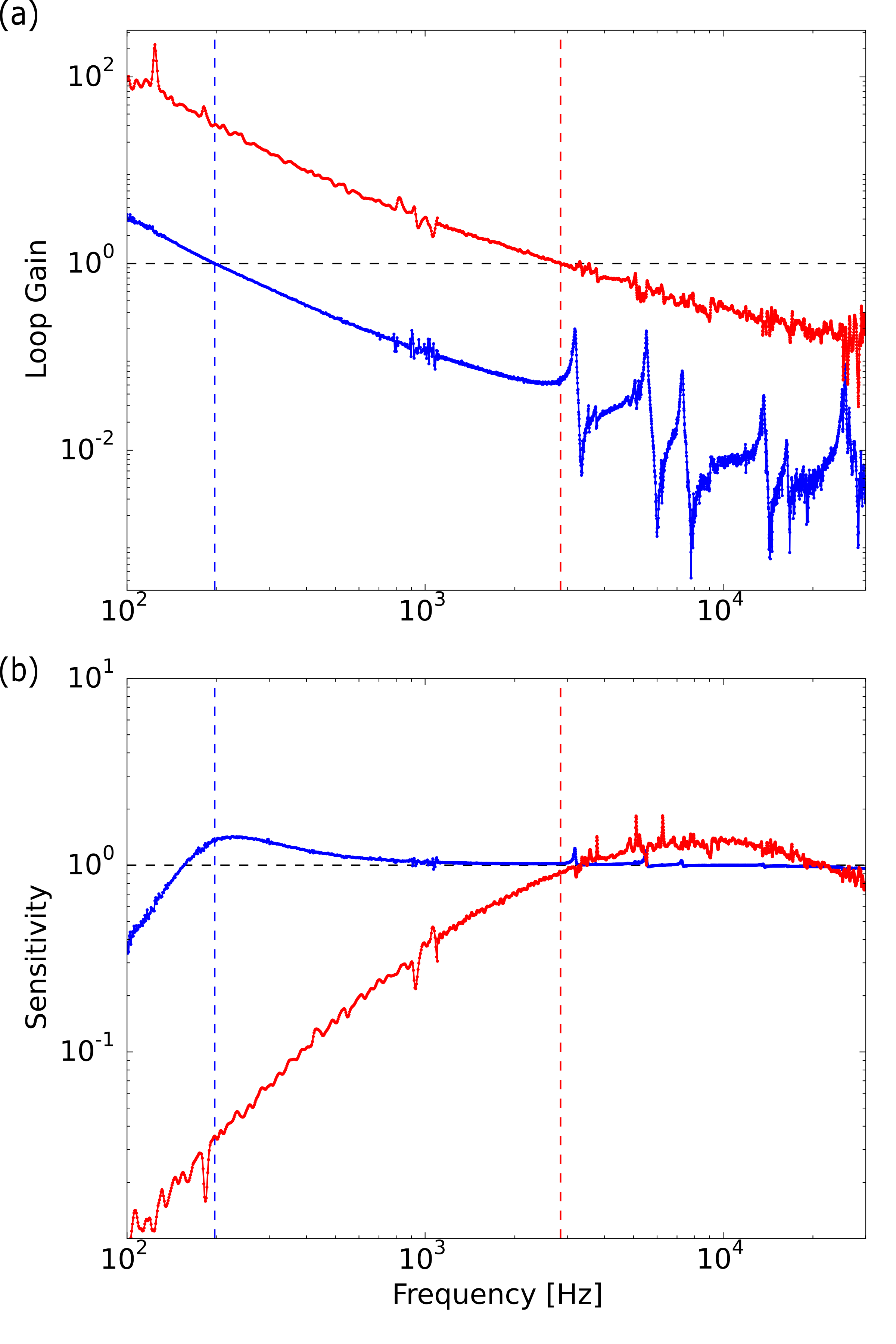}
\caption{\label{Figure:loop} \textbf{Loop gain and sensitivity}. (a) Magnitude of $KGF$ before (blue) and after (red) implementing the digital FIR filter to compensate the open-loop gain profile, and increasing the total gain. (b) Magnitude of the sensitivity (noise suppression factor) 1/(1+$KGF$) before (blue) and after (red) digital flattening and gain increase. The sensitivity in the uncompensated system displays small peaks at each mechanical resonance; further increasing the total gain, without compensating for the mechanical resonances, would result in noise amplification, and ultimately oscillation, at these resonant frequencies. In both plots, the horizontal dashed line indicates unity gain, and the vertical dashed line indicates the bandwidth (unity gain frequency), which has been enhanced by more than an order of magnitude. The final gain and phase margins are 4 and $70^{\circ}$, respectively.}
\end{figure}

\subsection{Loop gain and sensitivity}
Having cancelled the acoustical resonances with the FIR filter, we increase the total gain in the feedback system and hence the noise-suppression- and control- bandwidth. We record the ratio of the filtered output $y$ to the error signal $e$ while modulating the reference signal $r$; then, the loop gain $KGF = y/e$, and the sensitivity $1/(1+KGF) = e/r$ (see supplementary information for the calculation of the sensitivity).

Figure \ref{Figure:loop} shows the magnitude of (a) the loop gain and (b) the sensitivity before (blue) and after (red) implementing the FIR filter and increasing the total gain. Note that the unity-gain bandwidth has been increased by more than an order of magnitude from 200 Hz to 2.8 kHz, marked by the vertical dashed lines.

Figure \ref{Figure:loop}b shows correspondingly greater noise suppression for higher bandwidth.  The suppression occurs up to the bandwidth and then saturates at unity, where a servo bump is observed. For the uncompensated configuration, the strong servo bump is the result of a mismatch between the corner frequencies of $K (\omega = 2\pi \times 200$ Hz) and $GF (\omega = 2\pi \times 50$ Hz), leading to excess phase at the unity gain.

The maximum compensated bandwidth of 2.8 kHz is set by the presence of additional resonances beyond the Nyquist frequency of the FPGA FIR filter, including the piezo/mirror resonances (see supplementary information). The final gain and phase margins are 4 and $70^{\circ}$, respectively, indicating a robust lock \cite{mosk2005tutorial}.

\section{\label{sec:Conclusion}Conclusion}
We have demonstrated that a low-latency FPGA-based FIR filter is an extremely versatile tool for implementing sophisticated control schemes. With a sampling rate of $\sim$250 kHz, a memory time $\sim$0.1 seconds, and a delay of $\sim$2.6 $\mu$s, the filter is capable of canceling numerous acoustical resonances, thereby enhancing mechanical noise suppression far beyond what is practical in the analog regime. In addition to demonstrating the requisite FPGA FIR architecture, we introduce a minimal characterization scheme for the mechanical system that provides all parameters of the FIR filter in a single shot. We employ this approach to increase the feedback bandwidth of a locked optical resonator by an order of magnitude. 

Our approach is broadly applicable to any field that employs actively stabilized mechanical systems, including atomic microscopy, where it could be used to enhance the stability of the scanning tip \cite{binnig1985scanning, binnig1986atomic} as an alternative to sophisticated flexures \cite{kenton2011compact}, and gravitational wave detection, where it could suppress noise in the seismic stacks \cite{abbott2016observation,hardham2005quiet, matichard2015seismic}. The technique is directly extensible to higher frequency low-latency loop-shaping, at the expense of suppression of only lower Q features, or substantially larger FPGA's. The flexibility of this tool further suggests possibilities in control beyond the delay-limited bandwidth, to provide narrowband, high-frequency noise suppression.

\section{\label{sec:Acknowledgements}Acknowledgements}
We thank Ariel Sommer for fruitful discussions. This work was supported by the U.S. AFOSR (grant FP053419-01-PR) for FPGA development and the U.S. DOE (grant FP054241-01-PR) for relevant theoretical modeling. A.R. thanks the NDSEG for support.

\bibliography{TheBib}{}


\begin{figure}
\includegraphics[width=86mm]{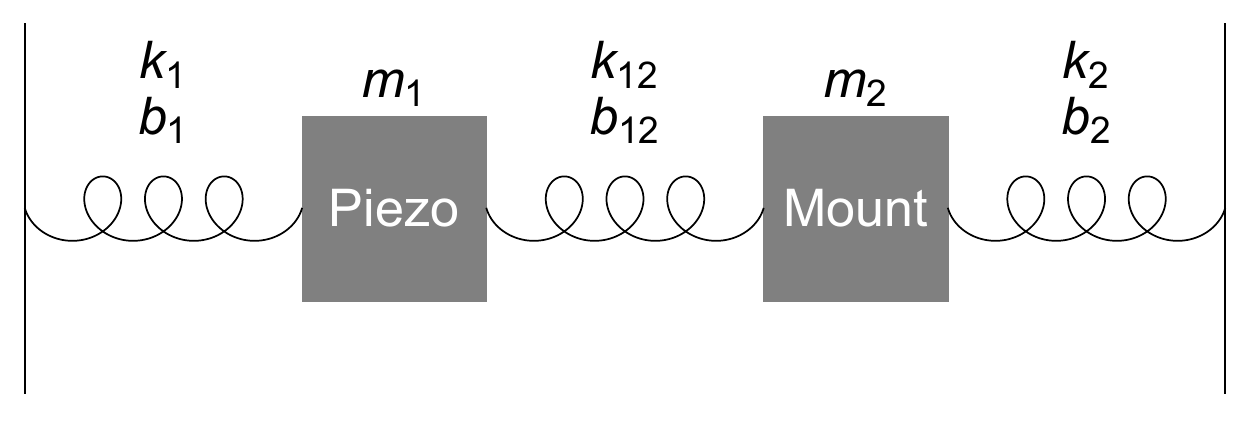}
\caption{\label{Figure:pole_zero_diagram} \textbf{Simple model: coupled damped harmonic oscillators.} Diagram of a model consisting of two coupled damped harmonic oscillators. Each oscillator $m_i$ is held to a rigid wall by a spring with spring constant $k_i$ and damping term $b_i$. The hybridization of the individual normal modes gives rise to a pole-zero pair and a separate pole in the transfer function of the piezo-actuator. The order in which the paired pole and zero appear depends on the relative strength between $k_{12}$ and $b_{12}$.}
\end{figure}

\begin{figure}
\includegraphics[width=86mm]{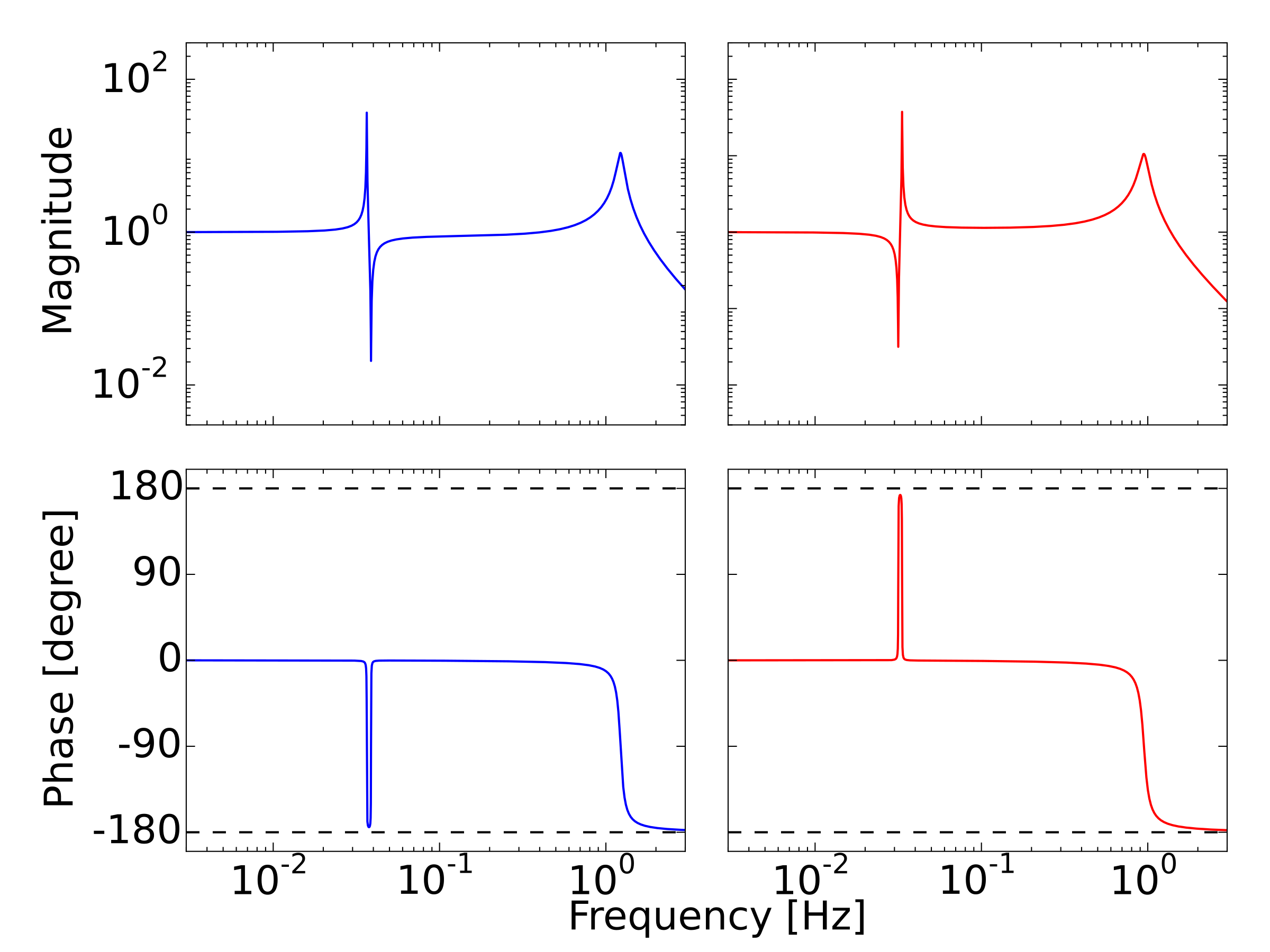}
\caption{\label{Figure:pole_zero_plot} \textbf{Model Transfer Function $T$.} (\emph{top}) Magnitude and \emph{(bottom}) Phase of the transfer function, versus Frequency. The interference between the piezo and the mount results in a pole-zero pair as well as a separate pole. (\emph{left}) For real coupling ($k_{12} \gg b_{12}$), the pole appears before the zero. (\emph{right}) For imaginary coupling ($k_{12} \ll b_{12}$), the zero appears before the pole. The poles and the zeros are accompanied by phase shifts of $-\pi$ and $\pi$, respectively.}
\end{figure}

\section{\label{sec:Supp}Supplementary Information}

\subsection{Physical origin of resonances}
The resonances (poles) and the anti-resonances (zeros) observed in the system transfer function arise from constructive and destructive interference between the normal mode of the piezo-actuator and each normal mode of the mounting structure. To see why this interference gives rise pole-zero pairs, we study a simple model consisting of two coupled damped harmonic oscillators, with masses $m_1$ and $m_2$, connected to rigid walls by damped springs $k_1$ ($b_1$) and $k_2$ ($b_2$) and to each other with a damped spring $k_{12}$ ($b_{12}$). See Figure \ref{Figure:pole_zero_diagram} for the diagram of the model.

The transfer function $T$ of the piezo-actuator is given by taking the first-row-first-column element of the $2 \times 2$ adjacency matrix:
\begin{equation}
T(\omega) = \left[\left(K + iB\omega - M\omega^2 \right)^{-1}\right]_{00}
\end{equation}
where $M=\left( \begin{smallmatrix}
	m_1 & 0 \\
	0 & m_2
\end{smallmatrix} \right)$, $B=\left( \begin{smallmatrix}
	b_1 & b_{12} \\
	b_{12} & b_2
\end{smallmatrix} \right)$, $K=\left( \begin{smallmatrix}
	k_1 & k_{12} \\
	k_{12} & k_2
\end{smallmatrix} \right)$, and the exponent denotes matrix inversion.

Figure \ref{Figure:pole_zero_plot} shows the magnitude and the phase of $T$, for two cases: (\emph{left}) $k_{12} \gg b_{12}$; (\emph{right}) $k_{12} \ll b_{12}$. In both cases, the transfer function exhibits a pole-zero pair (due to the piezo-mount coupling) as well as an isolated pole (due to the piezo alone); the order in which the pole and the zero appear in the pair depends on the relative strength of the spring ($k_{12}$) versus the damping ($b_{12}$) coupling. In the former case, the coupling is mostly through displacement, and the pole appears before the zero. In the latter case, the coupling is mostly through damping (providing an additional $\pi$ phase), and the zero appears before the pole. As expected, the poles and the zero are accompanied by phase shifts of $-\pi$ and $\pi$, respectively.

While it is in principle possible to fit the measured system transfer function $G$ with a physical model consisting of an arbitrary number of coupled damped harmonic oscillators, the mathematical model described in Section II.B offers significant advantages: First, the mathematical model eliminates extraneous degrees of freedom (which masses are coupled together and how strongly, how much and to whom they are damped, and which we drive and detect), and we only fit to poles and zeros we are interested in, namely those that limit the feedback bandwidth. Second, the fitting is much faster as the functional form does not require a matrix inversion.

\subsection{Measurement of the sensitivity}
In Section II.A., we derive and identify the sensitivity $1/(1+KGF)$ as the noise suppression factor by looking at the the block diagram (Figure \ref{Figure:blockdiagram}):
\begin{equation*}
x = \frac{KG}{1+KGF} r + \frac{1}{1+KGF} n
\end{equation*}

Experimentally, we cannot measure $n$ directly; instead, we measure $e$ while strongly modulating $r$. From the block diagram,
\begin{align}
e &= r - y \\
&= r - F(KGe + n) \\
e &= \frac{1}{1+KGF} r - \frac{F}{1+KGF} n
\end{align}
In this case, $r \gg n$ and the sensitivity is given by $e/r$. \\

\subsection{Temporal drift of open-loop response}
If the frequency dependence of the transfer function of the system drifts over time, the FIR filter function must be adjusted to compensate. For the application we have demonstrated, the acoustical resonances of the optical resonator are high Q, and can shift by as much as a linewidth as the piezo voltage (and hence its stiffness) changes in response to temperature variation of the piezo and the resonator structure \cite{waanders1991piezoelectric}.

We are working to resolve this first by stabilizing the piezo voltage by feeding back on the resonator temperature with heating coils, and second by simultaneously employing the FPGA as a network analyzer to monitor the slow drift of the resonances and adaptively re-calculating the inverse filter.

\subsection{Piezo-Actuator impedance}
\begin{figure}
\includegraphics[width=86mm]{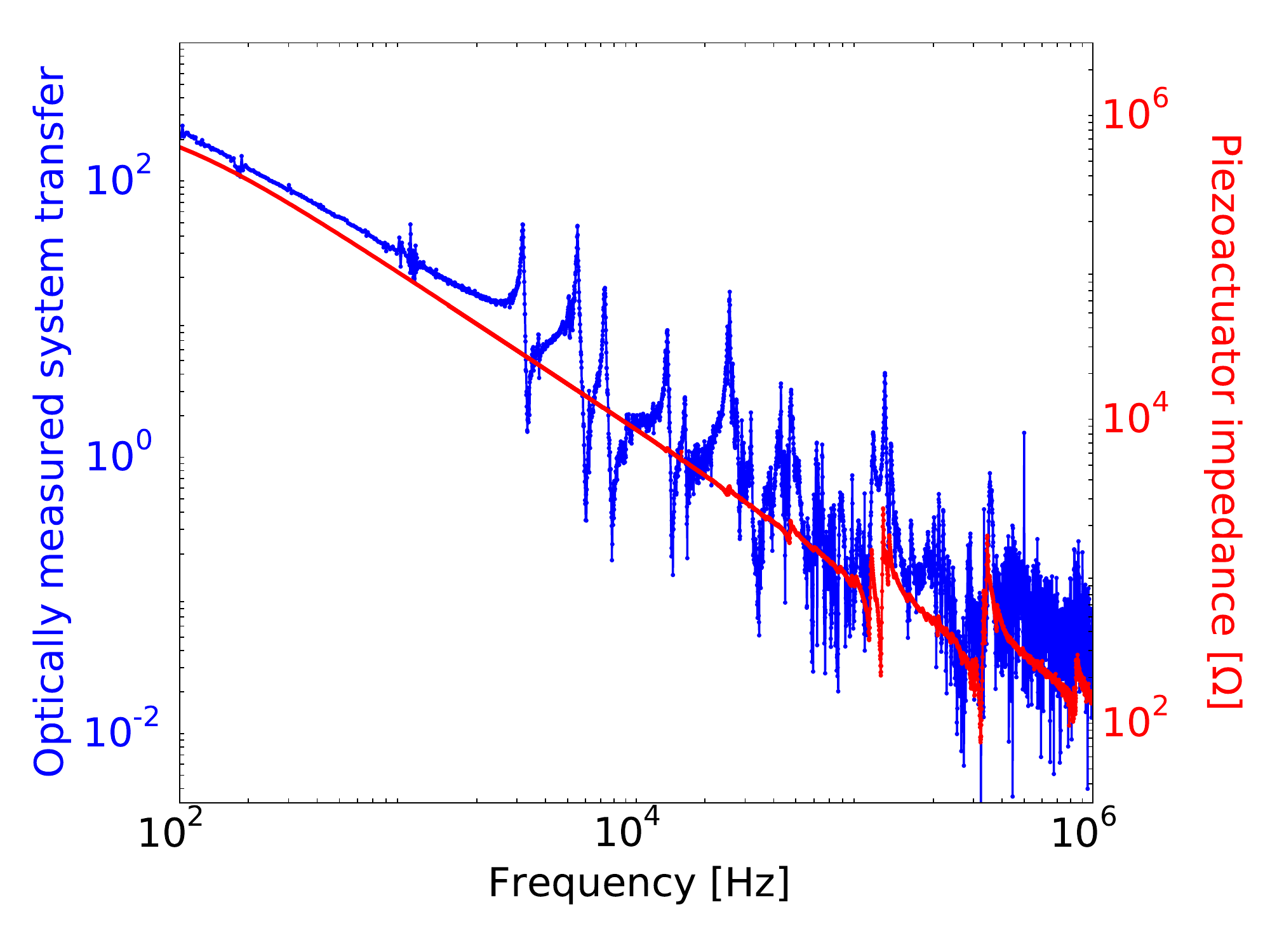}
\caption{\label{Figure:piezo_imp} \textbf{Piezoactuactor impedance} (red) plotted on top of the optically measured system transfer function (blue). The impedance strongly exhibits the resonances originating in the piezo-actuator itself and only weak coupling to the mount resonances.}
\end{figure}

The piezoelectric actuator is an electro-mechanical device: it expands or contracts depending on the applied voltage, and in turn, mechanical expansion and contraction of the piezo \emph{produces a voltage} that may itself be measured. As a consequence, the acoustical resonances can ``back act'' on the piezo-actuator, changing the frequency dependence of its electrical impedance. Compared to the optical interferometric measurements employed throughout this work, the electrical measurement is simpler to set up, as it does not require that the resonator be optically locked. The disadvantage is that the response is much weaker for resonances that do not originate in the piezo-actuator itself.  Nonetheless, comparison of the optically measured transfer function and the piezo-actuator impedance is a useful tool to reveal which of the acoustical resonances occur in the mount, and which in the piezo-actuator. See Figure \ref{Figure:piezo_imp} for the plot of the piezo-actuator impedance.

\subsection{Red Pitaya}
\begin{figure}
\includegraphics[width=86mm]{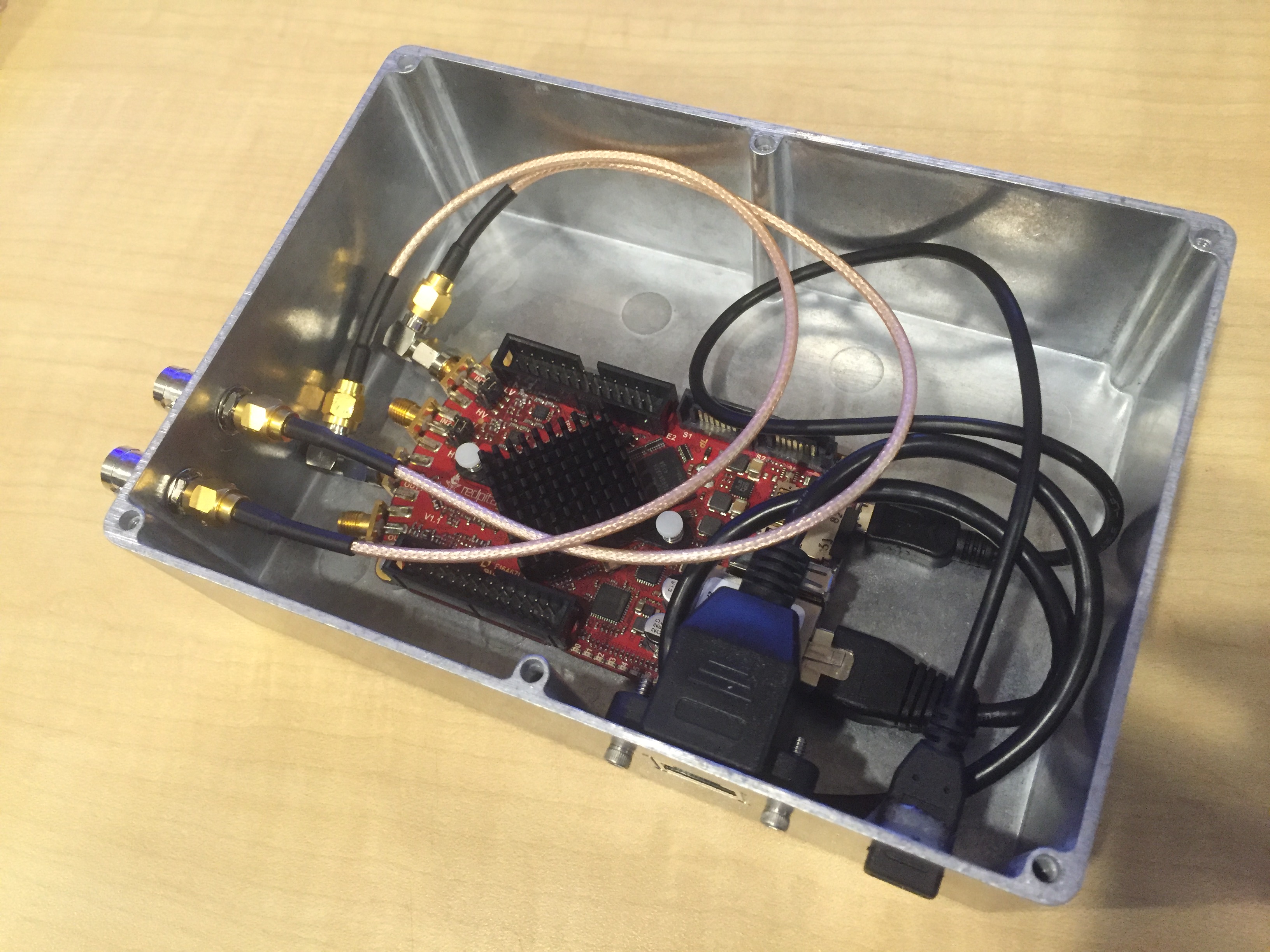}
\caption{\label{Figure:RPFig} \textbf{FPGA}. The FPGA is a Red Pitaya board, shown here in a custom housing.  On the left side are the input/output ports, for the raw and the modified output signals, $x$ and $y$, respectively.  On the right are the ports for an Ethernet cable and a power adapter.}
\end{figure}

See Figure \ref{Figure:RPFig} for a photograph of the Red Pitaya board.

\end{document}